\documentclass[epj]{webofc}
\usepackage[utf8]{inputenc}
\usepackage[varg]{txfonts}   
\usepackage{booktabs}
\usepackage{xcolor}
\definecolor{darkred}{rgb}{0.4,0.0,0.0}
\definecolor{darkgreen}{rgb}{0.0,0.4,0.0}
\definecolor{darkblue}{rgb}{0.0,0.0,0.4}
\usepackage[bookmarks,linktocpage,colorlinks,
    linkcolor = darkred,
    urlcolor  = darkblue,
    citecolor = darkgreen]{hyperref}
\usepackage{subfigure}
\wocname{EPJ Web of Conferences}
\woctitle{Lattice2017}
\begin{document}
%
\selectlanguage{english}
\title{%
Fighting topological freezing in the two-dimensional CP$^{N-1}$ model
\thanks{Talk given at the 35th International Symposium on Lattice Field Theory, 18 - 24 June 2017, Granada, Spain.}
}
\author{%
\firstname{Martin} \lastname{Hasenbusch}
}
\institute{%
Institut f\"ur Physik, Humboldt-Universit\"at zu Berlin,
Newtonstr. 15, 12489 Berlin, Germany
}
\abstract{%
We perform Monte Carlo simulations of the CP$^{N-1}$ model on the square lattice for $N=10$, $21$, and $41$. Our focus is on the severe slowing down related to instantons. To fight this problem we employ open boundary conditions as proposed by L\"uscher and Schaefer for lattice QCD. Furthermore we test the efficiency of parallel tempering of a line defect. Our results for open boundary conditions are consistent with the expectation that topological freezing is avoided, while autocorrelation times are still large. The results obtained with parallel tempering are encouraging.
}
\maketitle
\section{Introduction}\label{intro}

The CP$^{N-1}$ model shares fundamental properties such as asymptotic
freedom and confinement with QCD.
Therefore it serves as a toy model of QCD.
It has been shown \cite{DAdda:1978vbw,Witten79} 
that the model has a non-trivial vacuum 
structure with stable instanton solutions. It turned out that these
topological objects pose a particular problem in the simulation of the
lattice CP$^{N-1}$ model, similar to lattice QCD.

On the torus, in the continuum limit, the configuration space is decomposed into sectors
that are characterized by their topological charge. At finite lattice spacing, the
free energy barriers between such sectors increase as the lattice spacing decreases.
For Markov chain Monte Carlo algorithms that walk  in a quasi continuous fashion through
configuration space this means that they become essentially non-ergodic and slowing down
becomes dramatic. Numerical results are compatible with an increase of autocorrelation
times that is exponential in the inverse lattice spacing.
In the case of the CP$^{N-1}$ model this is numerically 
verified, for example, in refs. \cite{Campostrini,Vicari04,Flynnetal15}.
Modelling the autocorrelation times with a more conventional
power law Ansatz, large powers are needed to fit the data.
From a practical point of view, the consequence is that it becomes virtually impossible
to access lattice spacings below a certain threshold. The numerical studies show that in the case of the
CP$^{N-1}$ model the problem becomes worse with increasing $N$.
Since it is much less expensive to simulate the two-dimensional model than lattice QCD,
it is a good test bed for new ideas and algorithms that could overcome the severe
slowing down of the topological modes. For example simulated tempering \cite{MaPa92} has been studied
in ref. \cite{Vicari92} with moderate success. More recently,
``trivializing maps in the Hybrid Monte Carlo algorithm'' \cite{Engel:2011re}
or the ``Metadynamics'' method \cite{Metadynamics} have been tested.

A very principle solution of the problem had been suggested in ref. \cite{Luscher:2011kk}. By abandoning
periodic boundary conditions in one of the directions in favour of open ones, barriers between
the topological sectors are abolished.
The proposal has been further tested
\cite{Luscher:2012av,McGlMa14} and adopted in large scale simulations of lattice QCD with
dynamical fermions \cite{Bruno:2014ova,CLS}.
Here we shall probe in detail how open boundary conditions  effect the slowing
down  in the case of the CP$^{N-1}$ model. Since the CP$^{N-1}$ model is much cheaper
to simulate than lattice QCD, a larger range of lattice spacings can be studied and
autocorrelation functions can be computed more accurately.

Furthermore, we shall explore parallel tempering \cite{SW86,raex,Hukushima:1996xx,Earl:2005xx}
as a solution to our problem. Parallel tempering is a well
established approach in statistical
physics to overcome effective non-ergodicity due to a ragged free energy landscape.
The idea of parallel tempering  and similar methods
is to enlarge the configuration space such that the hills can be easily
by-passed.
A prototype problem is the study of spin-glasses, where parallel tempering
is mandatory. For recent work see for example ref. \cite{Janus2013}.
Typically a global parameter such as the temperature or
an external field is used as parameter of the tempering. Here instead, we shall discuss
a line defect. 

Finally we like to mention that for the CP$^{N-1}$ model dual formulations can be found. These can be
simulated by using the worm algorithm \cite{Prokofev:2001ddj,Ulli}. In these dual formulations there are no topological
sectors and hence severe slowing down does not occur in the simulation.

\section{The model}\label{model-1}
We consider a square lattice  with sites $x=(x_0,x_1)$, where 
$x_i \in \{0,1,2,...,L_i-1\}$. The lattice spacing is set to $a=1$.
This means that we trade a decreasing
lattice spacing for an increasing correlation length.
The action is 
\begin{equation}
 S = - \beta N \sum_{x,\mu}  \left(\bar{z}_{x+\hat \mu} z_x \lambda_{x,\mu} + z_{x+\hat \mu} \bar{z}_x 
 \bar{\lambda}_{x,\mu} -2  \right) \;,
\end{equation}
where $z_x$ is a complex $N$-component vector with
$z_x \bar{z}_x = 1$ and $\lambda_{x,\mu}$ is
a complex number with $\lambda_{x,\mu} \bar{\lambda}_{x,\mu} = 1$.
The gauge fields live on the links, which are denoted by $x, \mu$,
where $\mu \in \{0,1\}$ gives the direction
and $\hat \mu$ is a unit vector in $\mu$-direction. 
In $1$-direction we always consider periodic boundary conditions. In 
$0$-direction either open or periodic boundary conditions are considered.
We implement open boundary conditions in a crude way, simply setting 
$\beta=0$ for the links that connect $x_0=L_0-1$ and $x_0=0$.

\subsection{The observables}\label{obs-1}
We measure the energy, the magnetic susceptibility, the second moment 
and the exponential correlation length.  For the definition of these 
quantities see for example ref. \cite{Campostrini} or section II A of 
ref. \cite{MyCPN}.  Our main focus is on the topology of the field.
Motivated by eq.~(33) of ref. \cite{Campostrini}  we define the plaquette
angle
\begin{equation}
\theta_{plaq,x} =  \theta_{x,\mu} +  \theta_{x + \hat \mu,\nu}
 -\theta_{x + \hat \nu,\mu} - \theta_{x,\nu} - 2 n \pi \;\;\;,\;\;\;
\;\;\; \mu \ne \nu \;,
\end{equation}
where $\theta_{x,\mu}= \mbox{arg}\{\bar{z}_x z_{x+\hat \mu}\}$
and the integer $n$ is chosen such that $-\pi < \theta_{plaq,x} \le \pi$.
We define the topological charge density 
$q_x=\frac{1}{2 \pi}  \theta_{plaq,x}$. 
The topological charge on the lattice with periodic boundary conditions is
defined by
\begin{equation}
\label{Qplaq}
Q = \sum_x q_x = \frac{1}{2 \pi} \sum_x  \theta_{plaq,x} \;.
\end{equation}
The topological susceptibility is then given by
\begin{equation}
\label{chitper}
\chi_t = \frac{1}{V} \langle Q^2 \rangle  = \frac{1}{L_0 L_1} 
\left \langle \sum_{x y} q_x q_y  \right \rangle = \frac{1}{L_0 L_1}  
\left \langle \sum_{x_0 x_1} q_{x_0,x_1} \sum_{y_0 y_1} q_{y_0 y_1}  
\right \rangle = \frac{1}{L_0}  \left \langle \sum_{x_0} \tilde q_{x_0} 
 \sum_{y_0} \tilde q_{y_0}  \right \rangle \;,
\end{equation}
where we define $\tilde q_{x_0} = \frac{1}{\sqrt{L_1}} \sum_{x_1} q_{x_0,x_1}$. 
Note that the definition of the topological charge given in ref. \cite{Berg:1981er}
and (\ref{Qplaq}) are not equivalent at finite lattice spacing. We checked numerically that
the difference between the two definitions decreases quickly with increasing 
$\beta$. Also cooling of the configurations strongly reduces the difference.

On the lattice with periodic boundary conditions, $Q$ can take only integer
values. Naively, $\theta_{plaq,x}$ adds up to zero, since each link angle 
appears with both signs. A nontrivial result is due to the fact that   
$\theta_{plaq,x}$ is thrown back to the interval $[-\pi, \pi)$.

In the case of open boundary conditions, the definitions of susceptibilities 
have to be adapted. In order to avoid large finite size effects, the sites
with a distance less than $l_0$  from the open boundary  are not taken 
into account, when computing the observables.  Motivated by the rightmost 
part of eq.~(\ref{chitper}) we arrive at
\begin{equation}
\label{chitopen}
 \chi_{t,open} =  
\frac{1}{L_0-2 l_0}   \sum_{x_0 = l_0}^{L_0-l_0-1} \langle   \tilde q_{x_0}^2 \rangle \;
+2 \sum_{w=1}^{l_{max}}  \frac{1}{L_0-2 l_0 -w}  \sum_{x_0 = l_0}^{L_0-l_0-w-1} 
                    \langle   \tilde q_{x_0} \tilde q_{x_0+w} \rangle \;\;.
\end{equation}

\section{Basic algorithms}
As basic algorithm  we use a hybrid of the Metropolis, the heat bath
and the microcanonical overrelaxation algorithm.  To a large extend, we follow section III of ref. \cite{Campostrini}.
Let us first discuss the updates of the site variables and then the updates of the gauge fields.
In an elementary step of the algorithm  we update the variable at
a single site $x$, while keeping the gauge fields and the variables at all other 
sites fixed. The part of the action that depends on this site variable can be written as
as
\begin{equation}
 \tilde S(z_x) = - \mbox{Re} \; z_x \bar{F}_x \;,\;\; \mbox{where} \;\;
  F_x = 2 N \beta \; \sum_{\mu}
   \left [\bar{\lambda}_{x,\mu} z_{x+\hat \mu} +
          \lambda_{x-\hat \mu,\mu} z_{x-\hat \mu}  \right] \;.
\end{equation}
Note that the problem at this point is identical to the update of an
O$(2 N)$ invariant vector model with site variables of unit length. Instead
of $\bar{F}_x$ we would have to deal with the sum of the variables on the nearest neighbour
sites. The microcanonical update keeps $\tilde S(z_x)$ fixed, while the new
value of $z_x$ has maximal distance from the old one. It is given by eq.~(43a)
of ref. \cite{Campostrini}:
\begin{equation}
 z_x' = 2 \frac{ \mbox{Re} \; z_x \bar{F}_x}{|F_x|^2} F_x  - z_x \;.
\end{equation}
In addition to these updates, we have to perform updates that change the value
of the action. To this end we implemented a heat bath algorithm that is applied
to the subset of three of the $2 N$ components of $z_x$, where we count both 
the real and the imaginary parts. The heat bath update 
is identical to the one used in the simulation of the O(3)-Heisenberg model
on the lattice or for the update of SU$(2)$ subgroups in the simulation of
pure SU$(N)$ lattice gauge models \cite{CM82,Cr80}.
We run through all $N$ complex components
of $z_x$ taking the real and the imaginary part of the component as first two components for
the heat bath. The third component is randomly chosen among the real or imaginary parts of the
remaining $N-1$ components of $z_x$.
Note that the CPU-time required by the microcanonical overrelaxation update is about
one order of magnitude less than that for the heat bath update.

For fixed variables $z$ the gauge fields can be updated independently of each other. The action
reads
\begin{equation}
 \tilde S_g(\lambda_{x,\mu}) = - \mbox{Re} \lambda_{x,\mu} \bar{f}_{x,\mu} 
\;,\;\; \mbox{where} \;\;  f_{x,\mu} =  2 N \beta \; z_{x+\hat \mu} \bar{z}_x \;.
\end{equation}
Here we perform a four hit Metropolis update, where the stepsize was chosen such that the
acceptance rate is roughly $50\%$,
and a microcanonical update,
see eq.~(43b)  of \cite{Campostrini}.

\subsection{Autocorrelation times}
The performance of a Markov chain Monte Carlo algorithm is characterized
by the autocorrelation time. There are different definitions of the
autocorrelation time. These are based on the autocorrelation function.
The autocorrelation function of an estimator $A$ is given by
\begin{equation}
 \rho_A(t) = \frac{\langle A_{i} A_{i+t} \rangle - \langle  A \rangle^2}
                  {\langle A^2 \rangle - \langle  A \rangle^2} \;.
\end{equation}
The modulus of the autocorrelation function is bounded from above
by an exponentially decaying function. In practice one often finds that 
the autocorrelation function at large $t$ is given by  $\rho_A(t) \simeq 
c_A \exp(-t/\tau_{exp,A})$. 
The integrated autocorrelation time of the estimator $A$ is given by
\begin{equation}
\label{tauint}
 \tau_{int,A} = 0.5 + \sum_{t=1}^{\infty} \rho_A(t) \;.
\end{equation}
The summation in eq.~(\ref{tauint}) has to be truncated at some finite 
$t_{max}$. Since $\rho_A(t)$
is falling off exponentially at large distances, the relative statistical
becomes large at large distances. Therefore it is mandatory to truncate
the summation at some point that is typically much smaller than the
total length of the simulation.  In the literature one can find various
recommendations how this upper bound should be chosen. 
Fore example, Wolff \cite{Wolff:2003sm} proposes to
balance the statistical error with the systematic one that is due to the
truncation of the sum.

\section{Simulations with open boundary conditions}
In order to keep the fraction of discarded sites small, 
it seems useful to chose $L_0 \gg L_1$. On the other hand, since the 
time needed for topological objects to diffuse to the centre of the 
lattice or back to the boundary increases with increasing $L_0$, too 
large values of $L_0$ are not advisable. 
After performing preliminary simulations we decided to take $L_0=4 L_1$
throughout.  Furthermore we  take $l_0 \approx 10 \xi_{2nd}$ and $l_{max}=l_0$.
For $N=10$, using standard simulations and
periodic boundary conditions, $\xi_{2nd} \approx 23 $ can be reached 
\cite{Flynnetal15}. Hence it is hard to demonstrate a clear advantage
for open boundary conditions. Instead for $N=21$ it is virtually impossible 
to go beyond $\xi_{2nd} \approx 6$ by using periodic boundary conditions
and standard simulations. Therefore in the following we 
focus on  our simulations for $N=21$. We find that for 
$L_1 \gtrapprox 16 \xi_{2nd}$
finite $L_1$ effects can be ignored at the level of our statistical 
accuracy.  We performed simulations for a large number of $\beta$-values, 
ranging from $\beta=0.625$ up to $0.95$. For each value of $\beta$, we 
performed $2\times 10^6$ update cycles. For the larger values 
of $\beta$, we discarded $50000$ update cycles at the 
beginning of the simulation. One update cycle consists of one sweep over
all sites of the lattice using the heat bath algorithm, 
the 4 hit Metropolis update 
of the gauge fields, and finally $n_{ov}$ sweeps using the overrelaxation 
algorithm for both the site variables and the gauge fields.  We perform
a measurement of the observables for each cycle. Autocorrelation times 
are quoted in units of these update cycles.
The number of overrelaxation 
updates $n_{ov}$ is chosen to be proportional to the correlation length.
It increases from $n_{ov}=3$ for $\beta=0.625$  up to $n_{ov}=28$ for
$\beta=0.95$. The second moment correlation
length increases from $\xi_{2nd}=2.2968(5)$ at $\beta=0.625$ up to 
$18.2419(43)$ at $\beta=0.95$. 

\begin{figure}[thb]
  \centering
  \sidecaption
  \includegraphics[width=8.0cm,clip]{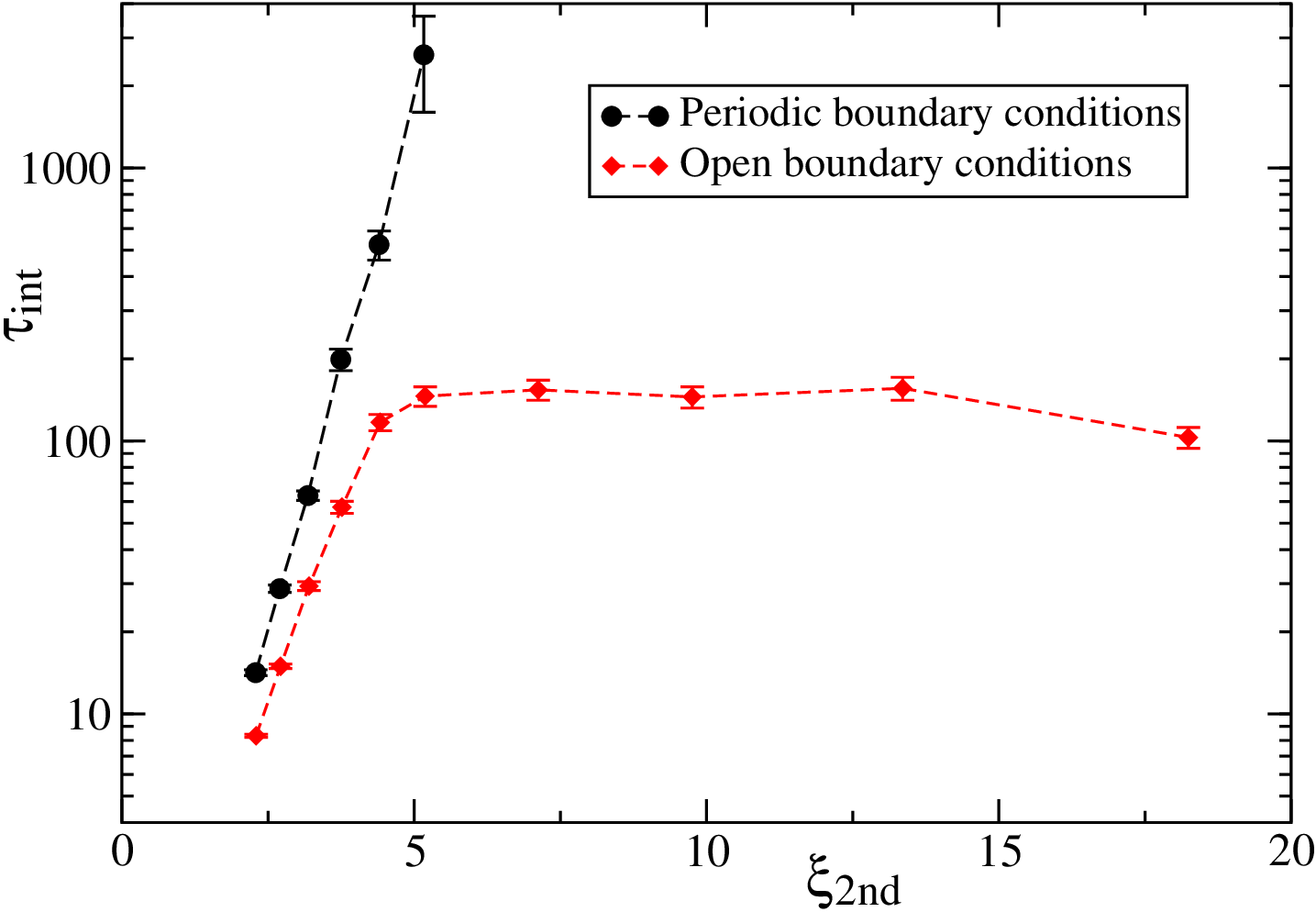}
  \caption{We plot the integrated autocorrelation time $\tau_{int}$ 
  of the topological
   susceptibility for both periodic and open boundary conditions as a function 
   of the second moment correlation length $\xi_{2nd}$ for $N=21$.
   Here $\tau_{int}$  is given in units of update cycles. 
   For large $\xi_{2nd}$ the effort for 
   one update cycle is proportional to the number of overrelaxation sweeps
   $n_{ov}$. Hence, the plateau of $\tau_{int}$ corresponds to a 
   dynamical critical exponent $z \approx 1$, taking single sweeps
   as unit of time.}
\label{fig:Tau21}
\end{figure}

Let us discuss the autocorrelation times of the topological 
susceptibility~(\ref{chitper},\ref{chitopen}).
For periodic boundary conditions, $\tau_{int}$ increases very rapid, 
compatible with exponential in the correlation length.
Instead, for open boundary conditions, we first see an increase that is
similar to that for periodic boundary conditions. The difference here
can be attributed to the different definitions of the topological 
susceptibility~(\ref{chitper},\ref{chitopen}).
Then, for $\xi_{2nd} \gtrapprox 5$ the autocorrelation time levels off
for open boundary conditions.

The behaviour in the case
of open boundary conditions can be explained along the lines of ref.
\cite{McGlMa14}. For $\xi_{2nd} \lessapprox 5$ changes of the topological
charge are dominantly due to the creation and destruction of instantons
in the bulk. Then for $\xi_{2nd} \gtrapprox 5$ the diffusion from and to
the boundaries completely dominates. This diffusion is not effected by
the severe slowing down. Our numerical results for $N=41$ confirm the 
conclusions drawn here for $N=21$. 

\section{Parallel tempering in a line defect}\label{Tempering}

\begin{figure}[thb]
  \centering
  \sidecaption
  \includegraphics[width=4.5cm,clip]{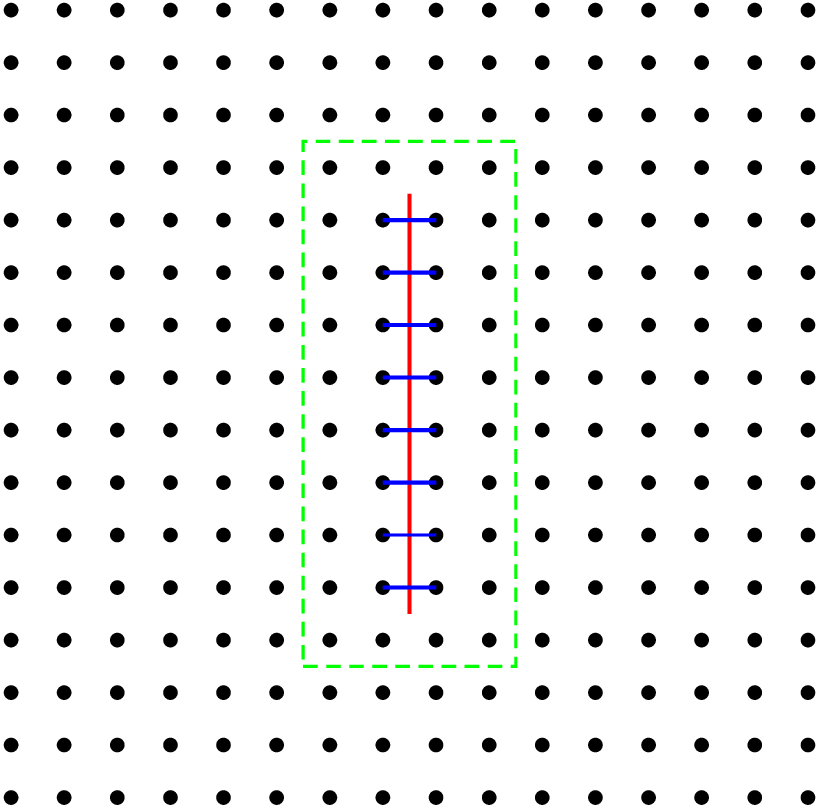}
  \caption{Sketch of a line defect. The red line indicates a line defect 
   of length $l_d$. The coupling on the blue links is reduced or, in the 
   extreme case, completely switched off. 
In our simulations we take for simplicity a linear interpolation: The couplings
on the blue links are multiplied by
$c_r(t)=1-t/(N_t-1)$. The homogeneous system corresponds to $t=0$, while for
$t=N_t-1$ the coupling along the defect line is completely eliminated.
   The green lines confine the 
   area of a rectangle centred around the defect line. 
   The size of the rectangle is $2 l_i \times (l_d + 2 l_i)$, where 
   $i$ gives the level. In our simulations $l_{i+1} = l_i/2$ for $i>1$.
    For each update cycle at level $i$ we perform $n_{i+1}$ update cycles 
    at level $i+1$.  At level $i=0$ we sweep over the whole lattice. 
    $l_1=2^m$, where $m$ is an integer and $l_1 \approx \xi$.  Furthermore
    $l_{i_{max}} =1$. 
}
  \label{fig-linedefect}
\end{figure}

In a parallel tempering simulation one introduces a sequence of $N_t$ systems 
that differ in one parameter of the action.
For each system there is a configuration $\{z,\lambda\}_t$.
The tempering parameter might have a physical meaning. In statistical physics 
simulations this parameter is mostly the temperature.
However it could  
also be a parameter that is introduced only for the sake of the simulation, as
it is the case here. At one end of the sequence there is the system that we 
want to study. In our case this is a lattice with $L_0=L_1$, periodic 
boundary conditions in both directions, and the coupling constant is the same for all links. For the system at the other end,
it should be easy to sample the whole configuration space.
Motivated by the success of the simulations with open boundary conditions,
we use a system with a line defect to this end. Such a line defect is 
sketched in fig. \ref{fig-linedefect}.  
For $l_d=L_1$ we recover open boundary conditions.

In addition to updates of the individual systems there are exchanges of 
configurations between the systems. 
A swap of configurations $\{z,\lambda \}_{t_1}'=\{z,\lambda \}_{t_2}$, 
$\{z,\lambda \}_{t_2}'=\{z,\lambda \}_{t_1}$ between $t_1$ and $t_2$  
is accepted with
the probability
\begin{equation}
\label{Aswap}
A_{swap}=\mbox{min}\left[1,\exp\left(-S_{t_2}(\{z,\lambda \}_{t_1}) 
                                     -S_{t_1}(\{z,\lambda \}_{t_2})
                                     +S_{t_1}(\{z,\lambda \}_{t_1})
                                     +S_{t_2}(\{z,\lambda \}_{t_2})   
  \right)
\right ] \;.
\end{equation}
In our simulations we run from $t_1=0$ up to $t_1=N_t-2$ in steps of one,
proposing to swap the configurations at $t_1$ and $t_2=t_1+1$. The number of
replica $N_t$ is chosen such that the acceptance rate for the swap of
configurations is larger than $30 \%$ for all $t_1$. Note that $S_{t_2}$ and
$S_{t_1}$ differ only on the defect line.
Typically the swap of configurations is alternating with standard updates
of the individual configurations. Typically, when the tempering parameter
is homogeneous in space, a sweep over the whole lattice is performed.
In contrast, here we temper in a defect that takes only a small fraction of
the lattice. Therefore it is advisable to update only some part
of the lattice that is centred around the defect.  To this end, we introduce
a sequence of rectangles of decreasing size, each associated with a level
of our update scheme. For details see 
Fig. \ref{fig-linedefect}. In one update cycle, at a given level,  we sweep
over the rectangle, updating all $N_t$ configurations: Once using the 
heat bath algorithm for the site variables and the 4 hit Metropolis update
for the gauge fields. Then follow $n_{ov,i}$ overrelaxation sweeps of the
site variables and the gauge fields. For small $i$, $n_{ov,i}$ is the same
as for our simulations with open boundary conditions. For larger $i$, smaller
values are taken. The update cycle at a 
given level is completed by a swap of configurations~(\ref{Aswap}).
We chose $n_{i}$ such that for each level 
of the update scheme, roughly the  same amount of CPU time is spent.
The larger $l_d$, the more topological objects can be generated or destroyed.
On the other hand, for increasing $l_d$,  $N_t$ has to be enlarged to keep 
the acceptance rate above $30 \%$. 
Our numerical study shows that $l_d \approx \xi$ is the optimal choice.
We perform a translation of the configuration for $t=0$ after each swap. 
This way changes in the topology are injected at any location on the lattice
and diffusion is not needed.  

We performed simulations
for $N=10$, $21$, and $41$ and various values of $\beta$. 
Let us discuss the simulation for $N=21$ and
$\beta=0.95$ in more detail. We used $l_d=16$ and $N_t=32$. 
The complete update cycle over all levels is characterized by
$n_1=24$, $n_2=n_3=n_4=3$, and $n_5=n_6=2$. The number of overrelaxation
updates per cycle is $28$, $14$, $7$, $7$, $3$, and $3$ at levels
$1$, $2$, $3$, $4$, $5$, and $6$, respectively.
We find that the acceptance rate is about $81.4 \%$
for the pair $t=0$ and $1$. It drops to  $39.4 \%$ for the pair
$t=26$ and $27$. Then it increases again to $47.3 \%$ for the pair $t=30$ and
$31$. 
The simulation, consisting of 50370 complete update cycles over all levels,  
took 25 days on
a 4 core PC running with 8 threads.  This is about the same CPU time
that is used for the corresponding run with open boundary conditions. The
error bar of the topological susceptibility is smaller by a factor of $2.3$
compared with the simulation with open boundary conditions.

\section{Physics results and comparison with the large $N$-expansion}\label{sec:physics}

\begin{figure}[tp]
   \centering
   \subfigure[
We plot the ratio $\xi_{exp}/\xi_{2nd}$ as a function of $\xi_{2nd}$.
]%
             {\includegraphics[width=0.50\textwidth,clip]{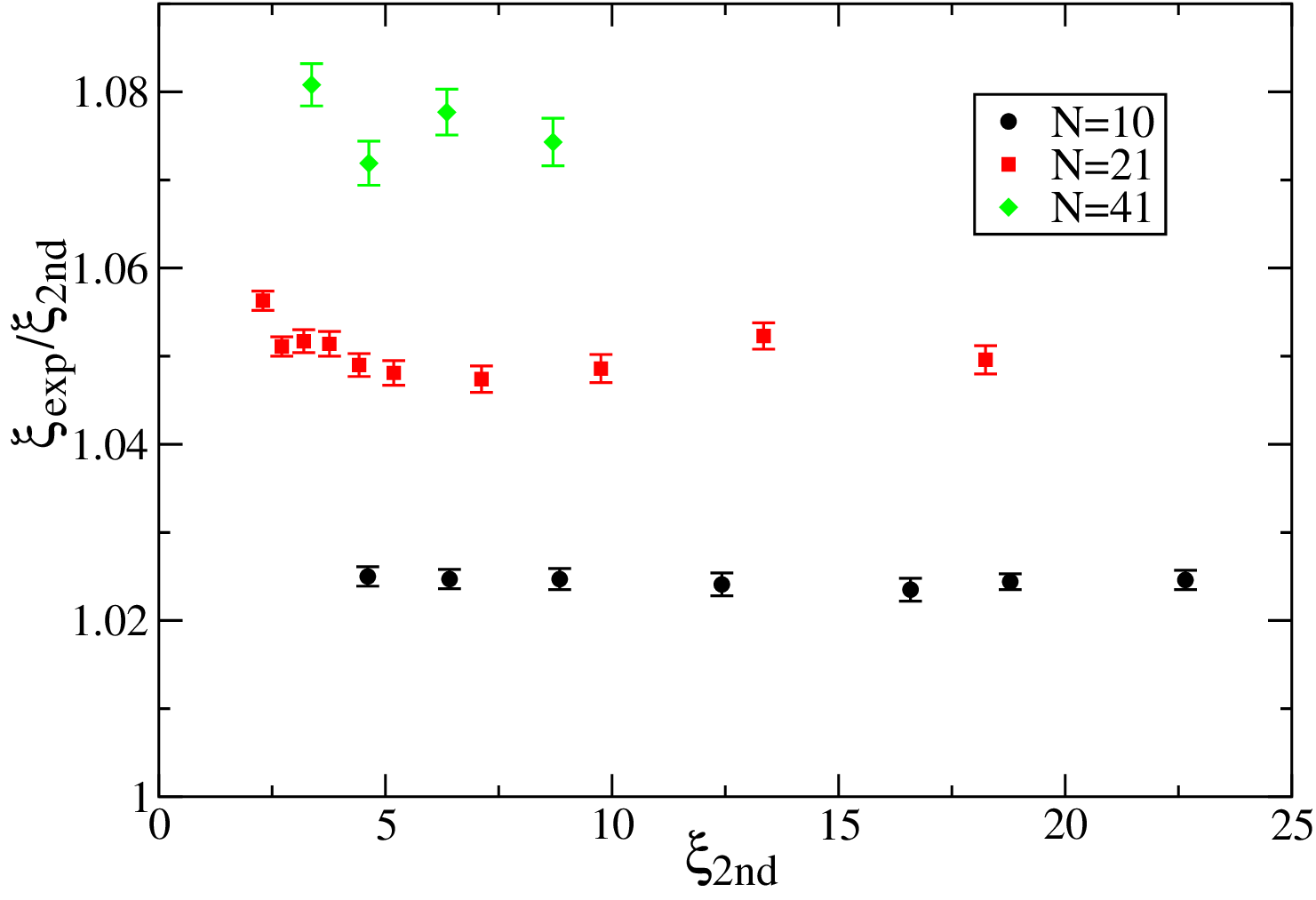}}\hfill
   \subfigure[ 
We plot $\xi_{2nd}^2 \chi_t$ as a function
of $\xi_{2nd}$.
]%
             {\includegraphics[width=0.50\textwidth,clip]{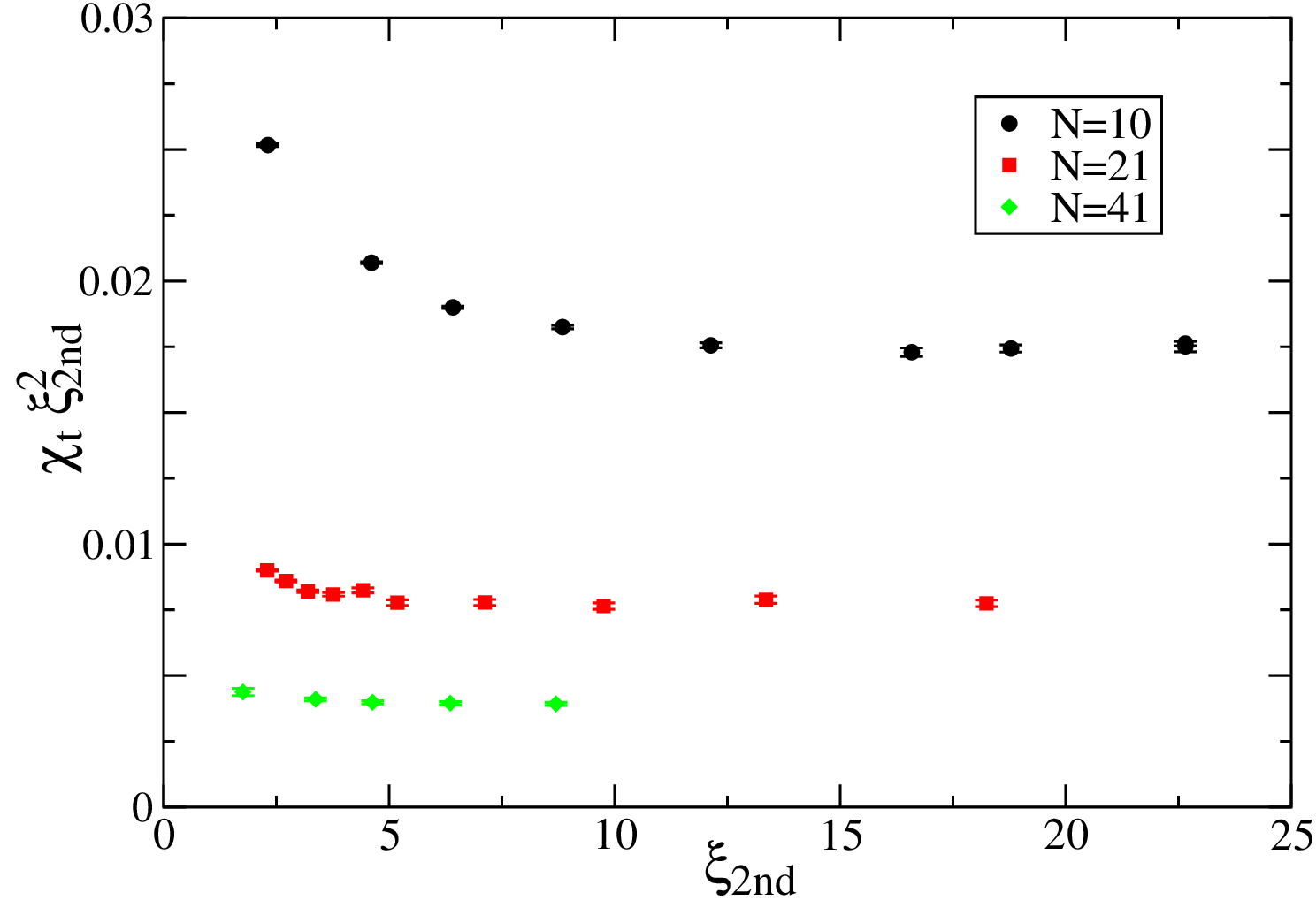}}\hfill
   \caption{Physics results for $N=10$, $21$ and $41$.
           }
   \label{fig:phys}
\end{figure}

Following ref. \cite{CaRo91}
\begin{equation}
 \frac{\xi_{2nd}}{\xi_{exp}} = \sqrt{\frac{2}{3}} + O\left(N^{-2/3}\right) \;.
\end{equation}
Our results obtained for $N=10$, $21$ and $41$, which are plotted in 
Fig. \ref{fig:phys} a are still quite far from
this asymptotic value. Therefore we abstain from estimating the coefficient of
the  $O(N^{-2/3})$ corrections.

The product $\chi_t \xi^2$ should  have a finite
continuum limit. For the exponential correlation length the $1/N$-expansion
gives \cite{ML78}
\begin{equation}
\chi_t \xi_{exp}^2 = \frac{3}{4 \pi N} + O\left(N^{-5/3}\right) \;\;.
\end{equation}

For the second moment correlation length a faster convergence with increasing
$N$ is obtained \cite{CaRo91}
\begin{equation}
\label{chitxi}
\chi_t \xi_{2nd}^2 = \frac{1}{2 \pi N} \left(1 - \frac{0.38088...}{N}  \right) +  O\left(N^{-3}\right) \;.
\end{equation}

In Fig. \ref{fig:phys} b we plot $\xi_{2nd}^2 \chi_t$ as a function
of $\xi_{2nd}$.  Looking at the figure, the numerical data seem to converge
nicely to the scaling limit.  Corrections to scaling seem to be smaller for
larger values of $N$.
Taking simply the largest values of $\beta$ for each $N$ we get
$\xi_{2nd}^2 \chi_t = 0.01737(8)$, $0.00767(5)$, and $0.00391(2)$ for
$N=10$, $21$, and $41$, respectively.
This can be compared with results quoted in the literature.
For $N=10$ one finds for example
$\xi_{2nd}^2 \chi_t = 0.01719(10)(3)$ and $0.0175(3)$ in
refs. \cite{Flynnetal15,Vicari04}, respectively. For $N=21$ one finds
$\xi_{2nd}^2 \chi_t =0.0080(2)$ and $0.0076(3)$  in refs.
\cite{Vicari04,Vicari92}, respectively. For $N=41$ we find in
ref. \cite{Vicari92} the results
$\xi_{2nd}^2 \chi_t = 0.0044(4)$ and $0.0036(4)$ for
$\beta=0.57$ and $0.6$, respectively. Our estimates are
essentially consistent with those presented in the literature.
In particular
for large values of $N$, we improved the accuracy of the estimates.
To see the effect of leading corrections, it is useful to multiply
$\xi_{2nd}^2 \chi_t$ by $2 \pi N$. Using our numbers, we get $1.091(5)$,
$1.012(7)$, and $1.007(5)$ for $N=10$, $21$, and $41$, respectively.
As already discussed in ref. \cite{Vicari04} it is a bit puzzling that
the numbers suggest a $1/N$ correction with the opposite sign as that of
eq.~(\ref{chitxi}).

\section{Summary and conclusions}
\label{summary}
We have shown that the severe slowing down in the simulation of the 
lattice CP$^{N-1}$ model can be avoided by using open boundary conditions
in one of the directions. We studied parallel tempering in a line defect 
as an alternative. Our numerical results are  encouraging. Focussing 
on the statistical error of the topological susceptibility,
the simulation with open boundary conditions is outperformed 
by a factor of about 4. The crucial question is, of course, 
whether parallel tempering in a defect structure is helpful in simulations
of lattice QCD.  A more detail account of this study is given in \cite{MyCPN}.

\section*{\small Acknowledgments}
I thank Stefan Schaefer for discussions.
This work was supported
 by the Deutsche Forschungsgemeinschaft (DFG) under the grant No HA 3150/4-1.

\clearpage

\end{document}